\documentclass[final,1p,times]{elsarticle}

\usepackage{graphicx}
\usepackage{amssymb}

\newcommand{\beq}{\begin{equation}}
\newcommand{\eeq}{\end{equation}}
\newcommand{\beqn}{\begin{eqnarray}}
\newcommand{\eeqn}{\end{eqnarray}}

\begin{document}

\begin{frontmatter}
  
\title{On the Period-Ratio-Mass-Ratio Correlation
of Extra-Solar Multi-Planet Systems}
\author{Sridhar Gajendran$^{a}$, 
Li-Chin Yeh$^{b}$, Ing-Guey Jiang$^{a}$
\footnote{e-mail: jiang@phys.nthu.edu.tw}
}

\address{
{$^{a}$Department of Physics and Institute of Astronomy,}\\
{National Tsing Hua University, Hsin-Chu, Taiwan}\\
{$^{b}$Institute of Computational and Modeling Science,}\\
{National Tsing Hua University, Hsin-Chu, Taiwan}
}

\begin{abstract} 
In order to gain possible hints for planet formation from the current data
of known extra-solar planets, 
the period-ratios and mass-ratios of adjacent planet pairs
in multi-planet systems are determined.
A moderate period-ratio-mass-ratio correlation is found
to have a correlation coefficient $r=0.5779$ with 99\% confidence interval 
(0.464, 0.672). In contrast, for non-adjacent planet pairs,
the correlation coefficient is $r=0.2820$ with 99\% confidence 
interval $(0.133, 0.419)$.
Our results reveal the imprint of planet-planet interactions 
of the adjacent planet pairs in a certain fraction of the multi-planet systems 
during the stage of planet formation. 
\end{abstract} 


\end{frontmatter}

\section{Introduction}

As the number of detected extra-solar planets (exoplanets) 
is increasing in a rapid pace, 
the field has become a topic of interest for nearly two decades. 
The physical properties and orbital configurations of 
the discovered exoplanets gives us a great insight 
into the formation of planetary systems, including our own Solar system. 
One popularly studied aspect of exoplanets is their distribution 
on the period-mass plane as the formation and evolution of planets 
are closely related to their orbital periods and masses.
Tabachnik \& Tremaine (2002) employed the maximum-likelihood method 
to determine the mass and period 
distributions of exoplanets and give an estimate of 
the fraction of stars which host exoplanets.
In addition, the continuity of planet masses and radii suggests that 
there is probably one major mechanism which governs planet formation. 
However, discoveries such as the ubiquitous presence of hot Jupiters
and super-earths seem to present challenges to the current 
understanding of planetary formation.

Contradicting models such as the core accretion and 
gravitational instability could both provide explanations 
for the presence of hot Jupiters.
Gravitational instability is found to work only at large distances 
from their host stars. For core accretion, the planet formation 
requires a pile-up of a huge amount of matter coming from a
larger distance range. 
Thus, the likely scenario is that the planets formed at 
greater distances and then migrated inwards (Raymond et al. 2006).
The final mass and location of planets therefore rely heavily on the 
interplay between its growth and migration.

A correlation is therefore expected between the orbital periods 
and masses of the planets. The first proposal for the existence 
of this correlation was made by Zucker \& Mazeh (2002).
This correlation could correspond directly to a deficit of massive planets 
with shorter orbital periods. This lack of close-in giant planets 
can be explained by tidal interactions (Patzold \& Rauer 2002)
between the planets and their central stars. 
These interactions would bring any planet within the Roche zone of the star, 
which would completely disrupt the formation of any close-in giant planets. 
This would also result in these stars having higher metallicities acquired 
from the absorbed planets.

On the other hand, the currently accepted core accretion model, 
which assumes orbital migration (Goldreich \& Tremaine 1980, Lin et al.  1996),
suggests that the present orbital radii of the planets are 
substantially smaller than the distance of their formation sites. 
The observed correlation might therefore reflect the correlation 
between the migration range of a planet and its mass. 
Generally, massive planets migrate slower (Nelson et al. 2000)
and are therefore left far away when the disk evaporates 
and this could also lead to the lack of giant planets with short periods.

The coupled period-mass functions and the period-mass correlation 
were studied simultaneously by Jiang et al. (2007) and Jiang et al. (2009).
Their study showed a positive correlation between 
the mass and period of the exoplanets. 
The planet population synthesis by Mordasini et al. (2009)
could give a theoretical explanation for this correlation. 
Their planet population synthesis used a simplified version of 
the extended core accretion model from Alibert et al. (2005)
as the formation model. Mordasini et al. (2009)
concluded that the abundance of gas giants with larger orbital periods 
in comparison to terrestrial planets, naturally leads to a correlation 
between the periods and masses of planets.

Furthermore, Mazeh \& Zucker (2003) discovered a surprisingly tight 
correlation of 0.9415 between the logarithms of period-ratio and mass-ratio
of adjacent planet pairs in multi-planet systems. 
However, this high correlation can be attributed 
to the lower number of samples, i.e. about 10, in their study. 
Jiang et al. (2015) re-examined this correlation with more samples 
and found a correlation of 0.5303. 

The interesting question is the effect of interaction 
between adjacent planets in multi-planet systems to this correlation. 
In fact, Laskar proposed a theory (Laskar 2000)
focusing on the interaction between adjacent planets 
in multi-planet systems using a planetesimal accretion model 
with the concept of angular momentum deficit (AMD). 
That model leads to a relation between period-ratio 
and mass-ratio of two adjacent planets in a multi-planet system
(Laskar \& Petit 2017; Yeh et al. 2020). 

Moreover, the above AMD model also leads to a planet spacing rule. 
In addition, Griv \& Gedalin (2005) discovered
that a theory of disk instability could lead to another spacing rule.  
With data of multi-planet systems, Bovaird \& Lineweaver (2013)
obtained a new two-parameter spacing relation empirically.
Huang \& Bakos (2014) later used Kepler data 
to search for possible new planets predicted by that relation,
and successfully identified five planetary candidates around
predicted positions. 

Inspired by the above development, we here 
re-visit the period-ratio-mass-ratio correlation (PRMR correlation). 
This is due to the fact that there are many newly confirmed
multi-planet systems during these five years, 
and thus the total number of multi-planet systems is nearly doubled. 
Moreover, Jiang et al. (2015) claimed a moderate PRMR correlation 
with correlation coefficient 0.5303. With a great fraction of newly added 
samples, it is very important to examine this PRMR correlation again
to investigate whether the correlation would be stronger or weaker.
In addition to presenting the PRMR correlation result 
of adjacent planet pairs, for the comparison purpose, 
we will also work out the results of non-adjacent pairs. 

Moreover, Jiang et al. (2015) did not give the exact list of
planets they employed in their study. This makes it difficult
to understand which exoplanets affect the results in a follow-up work.
We decide to provide the catalog of exoplanets employed in 
this work. This catalog will be very useful for
future statistical studies on any other possible correlation among 
these multi-planet systems.
 
Therefore, in this paper, we collect the updated publicly available 
data of mass and period of exoplanets in multi-planet systems. 
With these values of mass and period, 
we would give an updated study for the PRMR correlation.  
In \S 2, the data collection and analysis will be summarized.
In \S 3, our results of PRMR correlation will be presented.
The conclusion remarks will be in \S 4.

\section{The Method}

The exoplanet data of the multi-planet systems 
employed here is derived from the NASA exoplanet archive 
({\it http://exoplanetarchive.ipac.caltech.edu}). 
In order to obtain the period-ratio and mass-ratio of 
planet pairs from these multi-planet systems, 
only those systems which satisfy the following conditions are chosen:
\begin{enumerate}
	\item systems with at least two planets 
(as a minimum of two is required to determine the ratio);
	\item systems for which the mass and period values of all planets 
are completely known (i.e. systems which have at least one of these values 
missing are not taken into consideration).
\end{enumerate}
This leaves us with 228 multi-planet systems of useful data out of  
approximately 700 such systems. There are 563 exoplanets in
these 228 multi-planet systems and 
the planets' names, periods, masses are listed in Table 1.
Please note that, in this paper, the mass means the value of projected mass.
The period-ratio ($pr$) and mass-ratio 
($mr$) are calculated for any adjacent planet pairs in these systems. 
For example, for a system of three planets, the ratios are calculated between 
the intermediate one and innermost one and between the outermost 
one and the intermediate one. Thus, for a system of $N$ planets, 
we obtain $(N-1)$ sets of period-ratio ($pr=p_{i+1}/p_{i}$) 
and mass-ratio ($mr=m_{i+1}/m_{i}$)
between the adjacent $i-$th and $i+1$-th planets counted from the host star, 
where $p_i, p_{i+1}$ are the period of inner and outer planet,
$m_i, m_{i+1}$ are the mass of inner and outer planet, 
and $i=1, 2, 3, ..., N$.
The above calculations lead to 335 exoplanet pairs.

It is known that planets in resonant systems interact with each other 
significantly. The strong interaction between these planets in  
a resonant system would keep the period ratio between these planets 
to be special values as two-integer ratios (Snellgrove et al. 2001). 
Thus, the resonant pairs shall not be considered in our study of 
PRMR correlation. 

In order to find out if there are any pairs whose ratios are associated 
with resonances, a histogram of period ratios is plotted in Fig. 1(a).  
Since the maximum $pr$ value is more than 2500, the histogram is plotted 
only up to $pr$ = 10 in order to make the main part of the plot appear clear. 
The bin size considered here is 0.01 and the resonances occur around 
$pr$ = 1.5 (corresponds to 3:2 resonance) and another around $pr$ = 2 
(corresponds to 2:1 resonance). Fig. 1(b) and Fig. 1(c) reveal 
the presence of ratios associated with 3:2 and 2:1 resonances respectively. 
The planet pairs with these ratios are excluded from our sample set 
and are not considered for the PRMR correlation. 
Therefore, the number of useful planet pairs 
for calculating the period and mass ratios is reduced to 276.

\section{The Results}

Because the values of period and mass ratios cover a huge range, 
the natural logarithms of the period-ratio and mass-ratio are calculated 
and are denoted as $x$ and $y$, i.e. $x={\rm ln}(pr)$ and $y={\rm ln}(mr)$,  
in this paper. 
Fig. 2(a) shows the locations of all 276 samples 
on the $x$-$y$ plane. The $x$ and $y$ values present 
a positive correlation, and the least-square
best-fit line is $y = 0.5594 x - 0.3177$.

Following Press et al. (1992), 
the correlation coefficient between $x$ and $y$ is determined as
\begin{equation}
r = \frac{\sum(x_i - \overline{x})(y_i - \overline{y})}
{\sqrt{\sum(x_i - \overline{x})^2\ \sum(y_i - \overline{y})^2}},
\end{equation}
where $x_i$ is the individual $x$ value,  $\overline{x}$ is 
the mean of all $x$ values, 
$y_i$ is the individual $y$ value,  and $\overline{y}$ is 
the mean of all $y$ values. 

As suggested in Dowdy \& Wearden (1983),
the way to obtain the confidence intervals 
of correlation coefficients is to employ 
the Fisher's z-transformation. 
This is because the Fisher's z-transformation could lead to a 
normal distribution with known standard deviations.
After the correlation coefficient $r$ is obtained, 
a value $z$ is determined by (Press et al. 1992)
\begin{equation}
z = \frac{1}{2} ln \Bigg(\frac{1+r}{1-r}\Bigg).
\end{equation}
The standard deviation of this $z$ distribution is 
$\sigma = 1/\sqrt{n-3}$, where $n=276$, i.e. the number of samples.  

According to Table A.9 in the Appendix of Dowdy \& Wearden (1983),
for a normal distribution, 
the boundary of a 99\% confidence interval is at 2.58$\sigma$,
and the boundary of a 95\% confidence interval is at 1.96$\sigma$.
Thus, in the $z$-space, the 99\% confidence interval 
is $(z-2.58\sigma, z+2.58\sigma)$, and the 95\% confidence interval
is $(z-1.96\sigma, z+1.96\sigma)$.

Through the inverse Fisher's z-transformation,
\begin{equation}
r = \frac{e^{2z} - 1}{e^{2z} + 1},
\end{equation}
the above intervals can be transformed into the ones
in $r$-space easily.

Therefore,
the correlation coefficient of 276 samples in Fig. 2(a) 
is $r=0.5779$ with 99\% confidence interval $(0.464, 0.672)$.
Based on Cohen (1988), $0.1 < |r|\le 0.3$ means a weak correlation,
 $0.3 < |r|\le 0.5$ presents a moderate correlation,
and $0.5 < |r|\le 1.0$ indicates a strong correlation.
Our results show that there is a moderate correlation 
existing between period-ratio and mass-ratio of adjacent planet pairs 
among these multi-planet systems. 

For the purpose of comparison,  
as a next step, we here study the period-ratio and mass-ratio for 
two planets which may not necessarily be adjacent pairs. 
We randomly pick two planets from all 
563 exoplanets and calculate their period and mass ratios in a way
that the values of the planet with larger period are numerators. 
Because the chance that they happen to be a pair of adjacent planets
in the same multi-planet system is extremely small, 
they are regarded as non-adjacent pairs.  
To complete one simulation, 
the process is repeated until we have 276 non-adjacent pairs.
In order to do statistics, 10000 simulations are produced.

Fig. 2(b) shows the locations of all 276 non-adjacent samples 
on the $x$-$y$ plane of one typical simulation.
The $x$ and $y$ values present 
a weak positive correlation, 
and the least-square best-fit line is $y = 0.3832 x - 0.2602$.
The correlation coefficient  
is $r=0.2820$ with 99\% confidence interval $(0.133, 0.419)$.
This is a weak correlation. 

In order to see the distribution of correlation coefficients
of non-adjacent pairs in 10000 simulations, 
the histogram is presented in Fig. 3(a).
The peak and ranges of this distribution are consistent with 
a weak correlation.
Through the Fisher's z-transformation, 
Fig. 3(b) shows the distribution of $z$ values 
of non-adjacent pairs in 10000 simulations, where
the arrow indicates 
the corresponding $z$ value of adjacent planet pairs.
Therefore,
it is confirmed that the PRMR correlation of adjacent planet pairs
is much stronger than the PRMR correlation of non-adjacent planet pairs.

\section{Concluding Remarks}

In this paper, employing 276 adjacent planet pairs, 
we re-confirm a moderate PRMR correlation with a slightly larger
correlation coefficient $r=0.5779$ than the result in Jiang et al. (2015).
The 99\% confidence interval is (0.464, 0.672) and 
95\% confidence interval is (0.493, 0.651). 
Because this PRMR correlation is determined from adjacent planet pairs,
it indicates that two successive planets might have influenced each other 
during their formation. 

However, planet formation is a complicated process, which involves
interactions between many components, such as gas, dust grains, 
planetesimals, planetary embryos, and proto-planetary cores.
Moreover, while addressing their interactions, 
the growth, migration, and depletion of these components 
should be considered.

Assuming a phase that the gas is depleted and the disk 
is composed of planet embryos and planetesimals, 
Laskar (2000) and Laskar \& Petit (2017) developed a semi-analytic model 
which leads to a scaling relation between the 
period-ratio and mass-ratio of adjacent planets.
For a given power-law of disk surface-density distribution,
i.e. $R^{\beta}$, their model 
predicts that, for adjacent planet pairs, 
the natural logarithms of the period-ratio 
is proportional to the natural logarithms of the mass-ratio 
with a slope depending on the power-index $\beta$ only.

Considering that (a) probably only some planetary systems 
went through the phase
and processes suggested in Laskar \& Petit (2017),  
(b) the surface-density distributions of proto-planetary disks
could be different, and
(c) other mechanisms might need to be considered for 
some planetary systems, the moderate PRMR correlation 
we found in this paper could be a reasonable result. 
The examples for (c) could be the model in 
Christodoulou \& Kazanas (2019) for the TRAPPIST-1 planetary system
or the one suggested by
Griv \& Gedalin (2005) for the Solar system.

As for non-adjacent planet pairs, 
their PRMR correlation is found to be a weak one,
which could be related to a general trend of 
planet sizes during the formation in inner and outer part of 
planetary systems.
For example,  Zhu \& Wu (2018) found that when there is a super Earth 
in the inner part of a planetary system, the probability that
there is a cold Jupiter in the outer part of the same system is 
very high.

To conclude, our result of moderate PRMR correlation 
of adjacent planet pairs reveals the 
imprint of planet-plant interaction in some of the multi-planet systems.
In addition, 
the weak PRMR correlation of non-adjacent pairs 
could be related with a loose general 
trend of planet size distributions.

\section*{Acknowledgments}
We are grateful to the referee's suggestions.
This project is supported in part 
by the Ministry of Science and Technology, Taiwan, under
Ing-Guey Jiang's
Grant MOST 106-2112-M-007-006-MY3
and Li-Chin Yeh's 
Grant MOST 106-2115-M-007-014.

\clearpage
\begin{figure}[ht]
\includegraphics[width=1.0\textwidth]{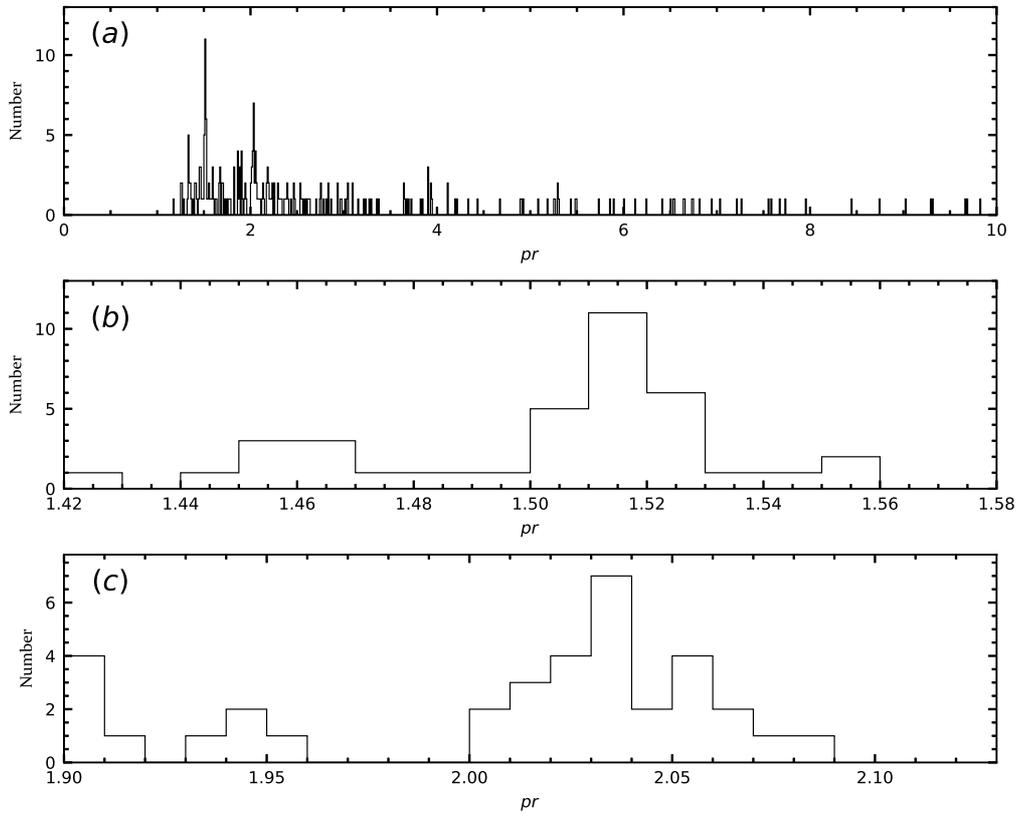}
\caption{(a)The histogram of period-ratio $pr$ 
upto $pr$ = 10.
(b)The histogram of period-ratio $pr$ near 3:2 resonance. 
(c)The histogram of period-ratio $pr$ near 2:1 resonance.
}
\label{fig1}
\end{figure} 

\clearpage 
\begin{figure}[ht]
\includegraphics[width=1.0\textwidth]{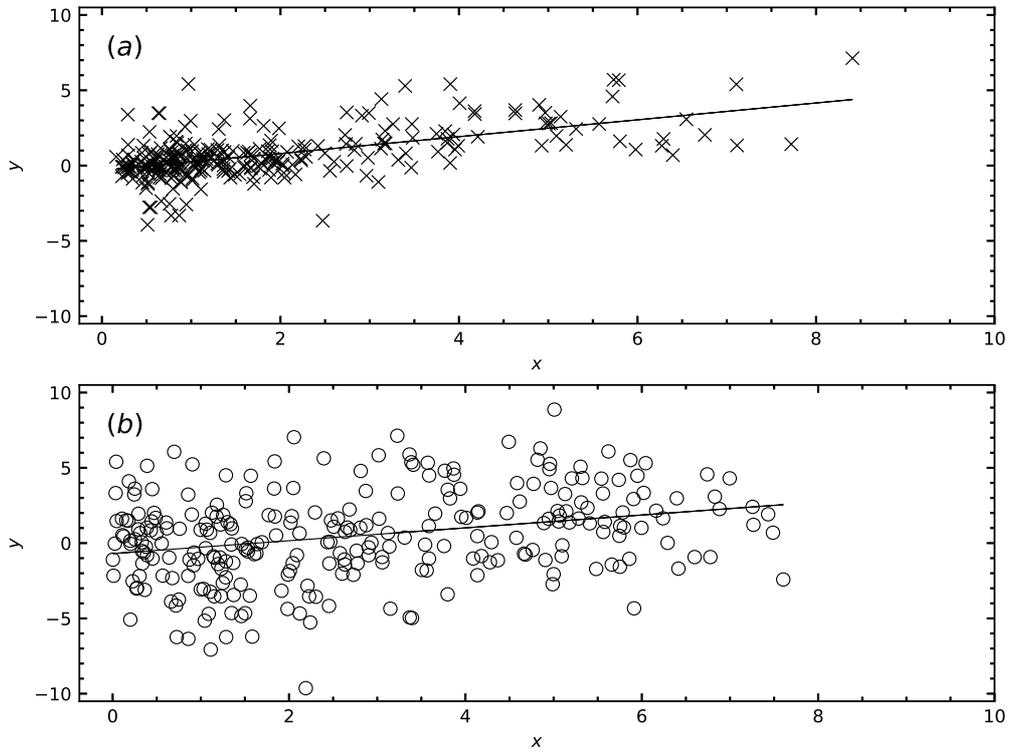}
\caption{(a)The distribution of 276 samples of adjacent pairs 
on the $x-y$ plane.
(b)The distribution of 276 samples of non-adjacent planet pairs
on the $x-y$ plane.
In both panels,
points are the locations of samples and the solid line is 
the best-fit line. 
}
\label{fig2}
\end{figure} 

\clearpage
\begin{figure}[ht]
\includegraphics[width=1.0\textwidth]{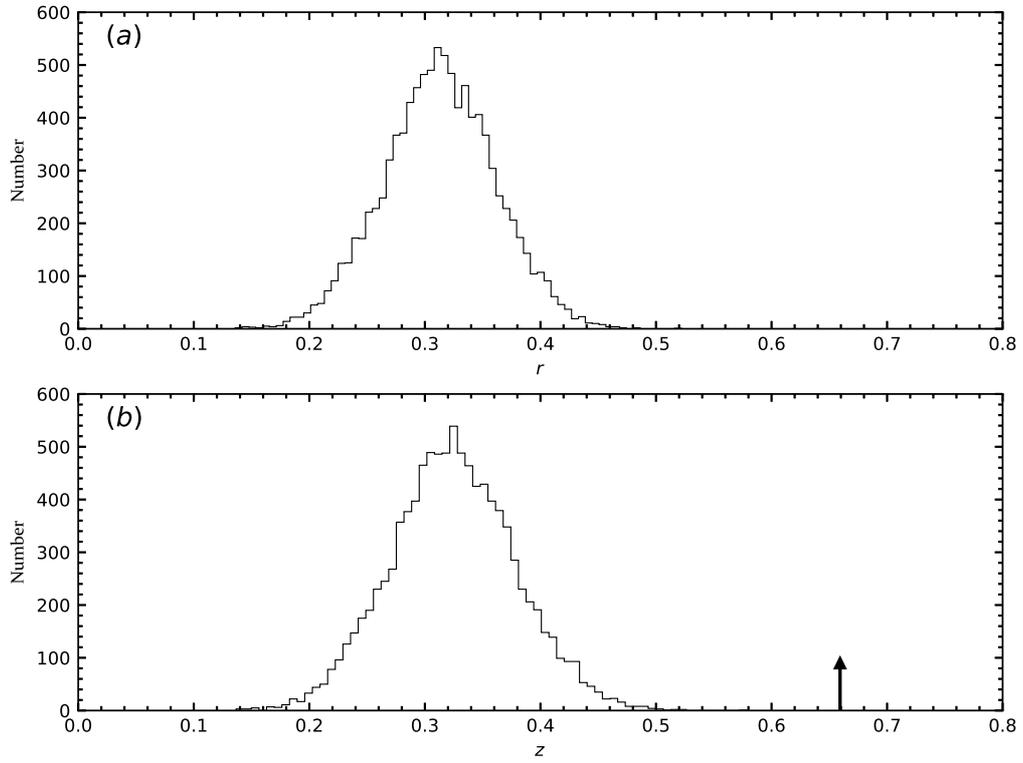}
\caption{(a)The histogram of correlation coefficients of 
non-adjacent planet pairs for 10000 simulations. 
(b)The corresponding $z$ histogram of non-adjacent planet pairs 
obtained using Fisher's z-transformation. The arrow indicates 
the corresponding $z$ value of adjacent planet pairs.}
\label{fig3}
\end{figure}

\clearpage
\begin{table*}
        \begin{tabular}{|l|l|l|l|l|}
                \hline
                ID & Planet Name & Period (day) & Planet Mass (Jupiter Mass) & References
                \\
                \hline
                
                1 &  24 Sex b  & 452.8      & 1.99       & Johnson et al. (2010) \\
                2 &  24 Sex c  & 883.0      & 0.86       & Johnson et al. (2010) \\
                3 &  47 UMa b  & 1078.0     & 2.53       & Gregory \& Fischer (2010)\\
                4 &  47 UMa c  & 2391.0     & 0.54       & Gregory \& Fischer (2010) \\
                5 &  47 UMa d  & 14002.0    & 1.64       & Gregory \& Fischer (2010) \\
                6 &  55 Cnc b  & 14.65152   & 0.8306     & Baluev (2015)\\
                7 &  55 Cnc c  & 44.4175    & 0.1714     & Baluev (2015) \\
                8 &  55 Cnc d  & 4825.0     & 3.878      & Baluev (2015) \\
                9 &  55 Cnc e  & 0.736539   & 0.02542    & Demory et al. (2016)  \\
                10 & 55 Cnc f  & 262.0      & 0.141      & Baluev (2015)          \\
                11 & 61 Vir b  & 4.215      & 0.016      & Vogt et al. (2010)     \\
                12 & 61 Vir c  & 38.021     & 0.057      & Vogt et al. (2010) \\
                13 & 61 Vir d  & 123.01     & 0.072      & Vogt et al. (2010)  \\
                14 & BD+20 2457 b  & 379.63  & 55.59     & Stassun et al. (2017) \\
                15 & BD+20 2457 c  & 621.99  & 12.47     & Niedzielski et al. (2009) \\
                16 & BD-06 1339 b  & 3.8728  & 0.027     & Lo Curto et al. (2013) \\
                17 & BD-06 1339 c  & 125.94  & 0.17      & Lo Curto et al. (2013) \\
                18 & BD-08 2823 b  & 5.6     & 0.04      & Stassun et al. (2010) \\
                19 & BD-08 2823 c  & 237.6   & 0.33      & Hebrard et al. (2010) \\
                20 & CoRoT-20 b    & 9.24285 & 4.3       & Rey et al. (2018) \\
                21 & CoRoT-20 c    & 1675.0  & 17.0      & Rey et al. (2018) \\
                22 & CoRoT-24 b    & 5.1134  & 0.018     & Alonso et al. (2014) \\
                23 & CoRoT-24 c    & 11.759  & 0.088     & Alonso et al. (2014) \\
                24 & CoRoT-7 b     & 0.85359163  & 0.01  & Stassun et al. (2017) \\
                25 & CoRoT-7 c     & 3.698   & 0.02643   & Queloz et al. (2009) \\
                26 & GJ 1132 b     & 1.628931 & 0.00522  & Bonfils et al. (2018) \\
                27 & GJ 1132 c     & 8.929   & 0.00831   & Bonfils et al. (2018) \\
                28 & GJ 1148 b     & 41.38   & 0.30425   & Trifonov et al. (2017) \\
                29 & GJ 1148 c     & 532.58  & 0.21414   & Trifonov et al. (2017) \\
                30 & GJ 163 b      & 8.63182 & 0.03335   & Bonfils et al. (2013) \\
                31 & GJ 163 c      & 25.63058 & 0.0214   & Bonfils et al. (2013) \\
                32 & GJ 163 d      & 603.95116  & 0.0925 & Bonfils et al. (2013) \\
                33 & GJ 273 b      & 18.6498 & 0.00909   & Astudillo-Defru et al. (2017) \\
                34 & GJ 273 c      & 4.7234  & 0.00371   & Astudillo-Defru et al. (2017) \\
                35 & GJ 3138 b     & 1.22003 & 0.0056    & Astudillo-Defru et al. (2017) \\
                36 & GJ 3138 c     & 5.974   & 0.01315  & Astudillo-Defru et al. (2017) \\
                37 & GJ 3138 d     & 257.8   & 0.03304  & Astudillo-Defru et al. (2017) \\
                38 & GJ 317 b      & 692.0   & 2.5      & Anglada-Escude et al. (2012) \\
                39 & GJ 317 c      & 5312.0  & 1.54     & Bryan et al. (2016) \\
                40 & GJ 3293 b     & 30.5987  & 0.07406  & Astudillo-Defru et al. (2017) \\
                41 & GJ 3293 c     & 122.6196 & 0.06636  & Astudillo-Defru et al. (2017) \\
                42 & GJ 3293 d     & 48.1345  & 0.02391  & Astudillo-Defru et al. (2017) \\
                43 & GJ 3293 e     & 13.2543  & 0.01032  & Astudillo-Defru et al. (2017) \\
                44 & GJ 3323 b     & 5.3636   & 0.00636  & Astudillo-Defru et al. (2017) \\
                45 & GJ 3323 c     & 40.54    & 0.00727  & Astudillo-Defru et al. (2017) \\
                46 & GJ 3998 b     & 2.64977  & 0.00777  & Affer et al. (2016) \\
                47 & GJ 3998 c     &  13.74   & 0.0197   & Affer et al. (2016) \\
                48 & GJ 581 b      & 5.3686   & 0.0497   & Robertson et al. (2014) \\
                49 & GJ 581 c      & 12.914   & 0.0173   & Robertson et al. (2014) \\
                50 & GJ 581 e      &  3.149   & 0.0053   & Robertson et al. (2014) \\       
                .. &   ...     & ...       & ...         & ...              \\
                \hline
        \end{tabular}
\caption{The Exoplanet Data. The ID, name, orbital period, projected mass,
          and reference are listed. Note that only a small portion 
          of the table is shown here.
       The full table of 563 planets is available in its entirety in 
       machine-readable form.}
\end{table*}


\begin{thebibliography}{plain}

\bibitem{A}
Alibert, Y., Mordasini, C., Benz, W., Winisdoerffer, C.,
Models of giant planet formation with migration and disc evolution,
2005, A\&A, 434, 343-353

\bibitem{Bo}
Bovaird, T., Lineweaver, C. H.,
Exoplanet predictions based on the generalized Titius–Bode relation,
2013, MNRAS, 435, 1126

\bibitem{Ch}
Christodoulou, D. M., Kazanas, D.,
Predicting additional planets in TRAPPIST-1,
2019,
Research Notes of the American Astronomical Society, 3, 50 

\bibitem{Co}
Cohen, J., Statistical power analysis for the behavioral
sciences, 1988,
Lawrence Erlbaum Associates, New Jersey

\bibitem{D}
Dowdy, S.,  Wearden, S., 
Statistics for research, 1983, John Wiley \& Sons Inc., New York

\bibitem{G}
Goldreich, P., Tremaine, S.,
Disk-satellite interactions,
1980, ApJ, 241, 425-441

\bibitem{Gr}
Griv, E., Gedalin, M.,
Formation of the solar system by instability, 
2005, 
Proceedings of IAU Colloquium 197: 
Dynamics of Populations of Planetary Systems,
Edited by Z. Knezevic and A. Milani, 
page 97, Cambridge University Press, Cambridge 

\bibitem{Hu}
Huang, C. X., Bakos, G. A.,
Testing the titius-bode law predictions for Kepler multiplanet systems,
2014, MNRAS, 442, 674

\bibitem{J07}
Jiang, I.-G., Yeh, L.-C., Chang, Y.-C., Hung, W.-L.,
On the mass-period distributions and correlations of extrasolar planets, 
2007, AJ, 134, 2061-2066

\bibitem{J09}
Jiang, I.-G., Yeh, L.-C., Chang, Y.-C., Hung, W.-L.,
Construction of coupled period-mass functions in extrasolar planets 
through a nonparametric approach, 
2009, AJ, 137, 329-336

\bibitem{J15}
Jiang, I.-G., Yeh, L.-C., Hung, W.-L., 
The period-ratio and mass-ratio correlation in 
extra-solar multiple planetary systems,
2015, MNRAS, 449, L65-L67

\bibitem{La2000}
Laskar, J.,
On the spacing of planetary systems,
2000, Phys. Rev. Lett., 84, 3240

\bibitem{LaskarPetit}
Laskar, J., Petit, A. C.,
AMD-stability and the classification of planetary systems,
2017, A\&A, 605, A72 

\bibitem{Li}
Lin, D. N. C., Bodenheimer, P., Richardson, D. C.,
Orbital migration of the planetary companion of 51 Pegasi 
to its present location,
1996, Nature, 380, 606-607

\bibitem{Ma}
Mazeh, T., Zucker, S., 
A possible correlation between mass ratio and period ratio 
in multiple planetary systems,
2003, ApJ, 590, L115-L117

\bibitem{Mo}
Mordasini, C., Alibert, Y., Benz, W., 
Extrasolar planet population synthesis,
2009, A\&A, 501, 1139-1160

\bibitem{N}
Nelson, R. P., Papaloizou, J. C. B., Masset, F., Kley, W.,
The migration and growth of protoplanets in protostellar discs,
2000, MNRAS, 318, 18-36 

\bibitem{P}
Patzold, M., Rauer, H.,
Where are the massive close-in extrasolar planets ?,
2002, ApJ, 568, L117-L120

\bibitem{Press}
Press, W. H. et al.,
Numerical Recipes in Fortran,
1992, Cambridge University Press, Cambridge, U.K.

\bibitem{RM}
Raymond, S. N., Mandell, A. M., Sigurdsson, S.,
Forming habitable worlds with giant planet migration,
2006, Science, 313, 1413-1416

\bibitem{S}
Snellgrove, M. D., Papaloizou, J. C. B., Nelson, R. P.,
On disc driven inward migration of resonantly coupled planets 
with application to the system around GJ876,
2001, A\&A, 374, 1092-1099

\bibitem{T}
Tabachnik, S., Tremaine, S.,
Maximum-likelihood method for estimating the mass and period 
distributions of extrasolar planets,
2002, MNRAS, 335, 151-158

\bibitem{Y}
Yeh, L.-C., Jiang, I.-G., Gajendran, S.,
On the scaling and spacing of extra-solar multi-planet systems, 
2020, Astrophysics and Space Science, 365, 186 
 
\bibitem{Zh}
Zhu, W., Wu, Y.,
The Super Earth-Cold Jupiter relations,
2018, AJ, 156, 92

\bibitem{Zu}
Zucker, S., Mazeh, T.,
On the mass-period correlation of the extrasolar planets,
2002, ApJ, 568, L113-L116

\end{thebibliography}
\end{document}